\documentstyle{article}
\advance\textheight by -\headsep\advance\textheight by -\headheight
\def\bbbr{{\rm I\!R}} %
\def\bbbc{{\mathchoice {\setbox0=\hbox{$\displaystyle\rm C$}\hbox{\hbox
to0pt{\kern0.4\wd0\vrule height0.9\ht0\hss}\box0}}
{\setbox0=\hbox{$\textstyle\rm C$}\hbox{\hbox
to0pt{\kern0.4\wd0\vrule height0.9\ht0\hss}\box0}}
{\setbox0=\hbox{$\scriptstyle\rm C$}\hbox{\hbox
to0pt{\kern0.4\wd0\vrule height0.9\ht0\hss}\box0}}
{\setbox0=\hbox{$\scriptscriptstyle\rm C$}\hbox{\hbox
to0pt{\kern0.4\wd0\vrule height0.9\ht0\hss}\box0}}}}
\newlength{\surusor}
\renewcommand\baselinestretch{1}
\def\<{\langle}
\def\>{\rangle}
\def\k{\kappa}
\def\im{{\rm i}}
\def\tr{{\rm Tr\,}}
\def\iH{{\cal H}}

\def\iV{{\cal W}}
\def\iM{{\cal M}}
\def\pont{\,\cdot \,}

\begin{document}

\title{
Entropy, von Neumann and the von Neumann entropy
{\footnote{This work was supported by the Hungarian National Foundation
for Scientific Research grant no. OTKA T 032662 and published in
{\it John von Neumann and the Foundations of Quantum Physics}, eds.
M. R\'edei and M. St\"oltzner, Kluwer, 2001.}}
}

\author{
{\it Dedicated to the memory of Alfred Wehrl} \\ \\
D{\'e}nes Petz{\footnote{Mathematical Institute,
Budapest University of Technology and Economics}}
}
\bigskip

\date{}

\maketitle

The highway of the development of entropy is marked by many
great names, for example, Clausius, Gibbs, Boltzmann, Szil\'ard, von Neumann,
Shannon, Jaynes, and several others. In this article the emphasis
is put on von Neumann and on quantum mechanics. The selection of
the subjects reflects the taste (and the knowledge) of the author
and it must be rather restrictive. In the past 50 years entropy
has broken out of thermodynamics and statistical mechanics and
invaded communication theory, ergodic theory and shown up in
mathematical statistics, social and life sciences. It is
practically impossible to present all of its features. The
favourite subjects of entropy is about macroscopic phenomena,
irreversibility and incomplete knowledge. In the strictly
mathematical sense entropy is related to the asymptotics of
probabilities or it is a kind of asymptotic behaviour of
probabilities.

This paper is organized as follows. After a short introduction
to entropy, von Neumann's gedanken experiment is repeated, which
led him to the formula of thermodynamic entropy of a statistical
operator. In the analysis of his ideas we stress the role of (the
lack of) superselection sectors and summarize von Neumann's knowledge
about quantum mechanical entropy. The final part of this article
is devoted to some important developments of the von Neumann
entropy which were discovered long after von Neumann's work.
Subadditivity and interpretation of the von Neumann entropy as the
capacity of a communication channel are among those.

\section{General introduction to entropy}

The word ``entropy'' was created by Rudolf Clausius and it appeared
in his work ``Abhandlungen \"uber die mechanische W\"armetheorie''
published in 1864. The word has a Greek origin, its first part
reminds us of ``energy'' and the second part is from ``tropos'' which
means turning point. Clausius' work is the foundationstone of
classical thermodynamics. According to Clausius, the change of
entropy of a system is obtained by adding the small portions of
heat quantity received by the system divided by the absolute
temperature during the heat absorption. This definition is
satisfactory from a mathematical point of view and
gives nothing other than an integral in precise mathematical
terms. Clausius postulated that the entropy of a closed system
cannot decrease, which is generally referred to as the second law
of thermodynamics. On the other hand, he did not provide any
heuristic argument to support the law. This fact might partly
be responsible for the mystery surrounding entropy for a long time.
As an extreme, we can cite Alfred Wehrl who had the opinion in
1978 that ``the second law of thermodynamics does not appear to
be fully understood yet'' \cite{We}.

The concept of entropy was really clarified by Ludwig Boltzmann.
His scientific program was to deal with the mechanical theory of
heat in connection with probabilities. Assume that a macroscopic
system consists of a large number of microscopic ones, we simply
call them particles. Since we have ideas of quantum mechanics in
mind, we assume that each of the particles is in one of the
energy levels $E_1<E_2<\dots <E_m$. The number of particles in
the level $E_i$ is $N_i$, so $\sum_i N_i=N$ is the total number
of particles. A macrostate of our system is given by the
occupation numbers $N_1,N_2,\dots, N_m$. The energy of a
macrostate is $E=\sum_i N_i E_i$. A given macrostate
can be realized by many configurations of the $N$ particles,
each of them at a certain energy level $E_i$. Those
configurations are called microstates. Many microstates realize
the same macrostate. We count the number of ways of arranging
$N$ particles in $m$ boxes (i.e., energy levels) such that each
box has $N_1, N_2,\dots, N_m$ particles. There are
\begin{equation}
\left(\begin{array}{c}
N \\
 N_1,N_2,\cdots,N_m
\end{array} \right):= \frac{N!}{N_1!\,
N_2! \dots N_m!} \label{1}
\end{equation}
such ways. This multinomial coefficient is the number of
microstates realizing the macrostate $(N_1,N_2,\dots,N_m)$
and it is proportional to the probability of the macrostate if
all configurations are assumed to be equally likely. Boltzmann
called (\ref{1}) the thermodynamical probability of the macrostate, in
German ``thermodynamische Wahrscheinlichkeit'', hence the letter
$W$ was used. Of course, Boltzmann argued in the framework of
classical mechanics and the discrete values of energy came from
an approximation procedure with ``energy cells''.

If we are  interested in the thermodynamic limit $N$ increasing to infinity, 
we use the relative numbers $p_i:=N_i/N$ to label a macrostate and, instead
of the total energy $E=\sum_i N_i E_i$, we consider the average
energy pro particle $E/N=\sum_i p_i E_i$. To find the most
probable macrostate, we wish to maximize  (\ref{1}) under a certain
constraint. The Stirling approximation of the factorials gives
\begin{equation}
{1 \over N} \log \left(\begin{array}{c}
N \\
 N_1,N_2,\cdots,N_m \end{array}\right)
=H(p_1,p_2,\dots,p_m)+O(N^{-1}\log N), \label{2}
\end{equation}
where
\begin{equation}
H(p_1,p_2,\dots,p_m):=\sum_i -p_i \log p_i. \label{3}
\end{equation}
If $N$ is large then the approximation (\ref{2}) yields that
instead of maximizing the quantity (\ref{1}) we can maximize (\ref{3}).
For example, maximizing (\ref{3}) under the constraint $\sum_i p_i E_i= e$, we
get
\begin{equation}
p_i={ e^{-\lambda E_i} \over \sum_j e^{-\lambda E_j} },
\label{4}
\end{equation}
where the constant $\lambda$ is the solution of the equation
$$
\sum_i E_i {e^{-\lambda E_i} \over \sum_j e^{-\lambda E_j}}=e.
$$
Note that the last equation has a unique solution if $E_1
< e < E_m$, and the distribution (\ref{4}) is known as the
discrete Maxwell-Boltzmann law today.

Let $p_1,p_2,\dots,p_n$ be the probabilities of different
outcomes of a random experiment. According to Shannon, the
expression (\ref{1}) is a measure of our ignorance prior to the
experiment. Hence it is also the amount of information gained by
performing the experiment. (\ref{1}) is maximum when all the
$p_i$'s
are equal. In information theory logarithms with base 2 are used
and the unit of information is called bit (from binary digit).
As will be seen below, an extra factor equal to Boltzmann's
constant is included in the physical definition of entropy.

\section{Von Neumann's contribution to entropy}

The comprehensive mathematical formalism of quantum mechanics was
first presented in the famous book ``Mathematische Grundlagen der
Quantenmechanik'' published in 1932 by Johann von Neumann. In the
traditional approach to quantum mechanics, a physical system is
described in a Hilbert space: Observables correspond to selfadjoint
operators and statistical operators are asssociated with the
states. In fact, a statistical operator describes a
mixture of pure states. Pure states are the really physical
states and they are given by rank one statistical operators, or
equivalently by rays of the Hilbert space.

Von Neumann associated an entropy quantity to a statistical
operator in 1927 \cite{vN1} and the discussion was extended in his
book \cite{vN2}. His argument was a gedanken experiment on the ground
of phenomenological thermodynamics. Let us consider a gas of $N(\gg 1)$
molecules in a rectangular box $K$. Suppose that the gas behaves
like a quantum system and is described by a statistical operator
$D$ which is a mixture $\lambda |\varphi_1\>\< \varphi_1|
+(1-\lambda)|\varphi_1\>\<\varphi_2|$, $|\varphi_i\>\equiv \varphi$ 
is a state vector $(i=1,2)$. We may take $\lambda N$ molecules in
the pure state $\varphi _1$ and $(1-\lambda )N$ molecules in the
pure state $\varphi _2$. On the basis of phenomenological
thermodymanics we assume that if $\varphi _1$ and $\varphi _2$ are
orthogonal, then there is a wall which is completely permeable for
the $\varphi_1$-molecules and isolating for the $\varphi_2$-molecules.
(In fact, von Neumann supplied an argument that such a wall
exists if and only if the state vectors are orthogonal.) We add
an equally large empty rectangular box $K'$ to the left of the box $K$
and we replace the common wall with two new walls. Wall (a),
the one to the left is impenetrable, whereas the one to the right,
wall (b), lets through the $\varphi_1$-molecules  but keeps back the
$\varphi_2$-molecules. We add a third wall (c) opposite to (b) which
is semipermeable, transparent for the $\varphi_2$-molecules and
impenetrable for the $\varphi_1$-ones. Then we push slowly (a) and (c)
to the left, maintaining their distance. During this process the
$\varphi_1$-molecules are pressed through (b) into $K'$ and
the $\varphi_2$-molecules diffuse through wall (c) and remain
in $K$. No work is done against the gas pressure, no heat is developed.
Replacing the walls (b) and (c) with a rigid absolutely
impenetrable wall and removing (a) we restore the boxes $K$ and $K'$ and
succeed in the separation of the $\varphi _1$-molecules from the
$\varphi_2$-ones without any work being done, without any temperature
change and without evolution of heat. The entropy of the original
$D$-gas (\, with density $N/V$\, ) must be the sum of
the entropies of the $\varphi_1$- and $\varphi_2$-gases
(~with densities $\lambda \,N/V$ and $(1-\lambda )N/V$,
respectively.~) If we compress the gases
in $K$ and $K'$ to the volumes $\lambda V$ and $(1-\lambda)V$,
respectively, keeping the temperature $T$ constant by means of a
heat reservoir, the entropy change amounts to $\k\lambda N\log \lambda$
and $\k(1-\lambda)N\log (1-\lambda )$, respectively. Indeed, we have
to add heat in the amount of $\lambda_i N \k T \log \lambda_i$~($<0$) when
the $\varphi_i$-gas is compressed, and dividing by the temperature
$T$ we get the change of entropy. Finally, mixing the
$\varphi_1$- and $\varphi_2$-gases of identical density we obtain a
$D$-gas of $N$ molecules in a volume $V$ at the original
temperature. If $S_0(\psi,N)$ denotes the entropy of a
$\psi$-gas of $N$ molecules (~in a volume $V$ and at the given
temperature~), we conclude that
\begin{eqnarray*}
\lefteqn{S_0(\varphi_1,\lambda N)+S_0(\varphi_2,(1-\lambda)N)} \\
&& =S_0(D,N)+\k\lambda
N\log\lambda +\k(1-\lambda)N\log (1-\lambda)
\end{eqnarray*}
must hold, where $\k$ is Boltzmann's constant. Assuming that
$S_0(\psi,N)$ is proportional to $N$ and dividing by $N$ we have
\begin{eqnarray}
\lefteqn{\lambda S(\varphi_1)+(1-\lambda)S(\varphi_2)} \nonumber \\
&& =S(D)+\k\lambda\log \lambda +\k(1-\lambda)\log (1-
\lambda)\, ,\label{5}
\end{eqnarray}
where $S$ is certain thermodynamical entropy quantity (~relative to
the fixed temperature and molecule density~). We arrived at the
mixing property of entropy, but we should not forget about the initial
assumption: $\varphi_1$ and $\varphi_2$ are supposed to be
orthogonal. Instead of a two-component mixture, von Neumann
operated by an infinite mixture, which does not make a big
difference,  and he concluded that
\begin{equation}
S\Big(\sum_i \lambda_i |\varphi_i\>\<\varphi_i|\Big)=
\sum_i \lambda_i S(|\varphi_i\>\<\varphi_i|)-\k \sum_i \lambda_i \log
\lambda_i. \label{6}
\end{equation}

Before we continue to follow his considerations, let us note that
von Neumann's argument does not require that the statistical operator
$D$ is a mixture of pure states. What we really needed is the
property $D=\lambda D_1 + (1-\lambda)D_2$ in such a way that
the possible mixed states $D_1$ and $D_2$ are disjoint. $D_1$ and
$D_2$ are disjoint in the thermodynamical sense, when there is a
wall which is completely permeable for the molecules of a
$D_1$-gas and isolating for the molecules of a $D_2$-gas. In
other words, if the mixed states $D_1$ and $D_2$ are disjoint,
then this should be demonstrated by a certain filter.
Mathematically, the disjointness of $D_1$ and $D_2$ is
expressed in the orthogonality of the eigenvectors
corresponding to nonzero eigenvalues of the two density matrices.
The essential point is in the remark that equation (\ref{5})
must hold
also in a more general situation when possibly the states do not
correspond to density matrices but orthogonality of the states
makes sense:
\begin{eqnarray}
\lefteqn{\lambda S(D_1)+(1-\lambda)S(D_2)}\nonumber \\
&& =S(D)+\k\lambda\log \lambda +\k(1-\lambda)\log (1-
\lambda)\, \label{7}
\end{eqnarray}

Equation (\ref{6}) reduces the determination of the
(thermodynamical)
entropy of a mixed state to that of pure states. The so-called
Schatten decomposition $\sum_i \lambda_i |\varphi_i\>\<\varphi_i|$
of a statistical operator is not unique even if $\<\varphi_i,
\varphi_j\>=0$ is assumed for $i \ne j$. When $\lambda_i$ is an
eigenvalue with multiplicity, then the corresponding eigenvectors
can be chosen in many ways. If we expect the entropy $S(D)$ to be
independent of the Schatten decomposition, then we are led to the
conclusion that $S(|\varphi\>\<\varphi|)$ must be independent of
the state vector $|\varphi\>$. This argument assumes that there
are no superselection sectors, that is, any vector of the Hilbert
space can be a state vector. On the other hand, von Neumann
wanted to avoid degeneracy of the spectrum of a statistical
operator (as well as the possible degeneracy of the spectrum of
observables as we shall see below).

Von Neumann's proof of the property that $S(|\varphi\>\<\varphi|)$
is independent of the state vector $|\varphi\>$ was different. He
did not want to refer to a unitary time development sending one
state vector to another, because that argument requires great
freedom in choosing the energy operator $H$. Namely, for any
$|\varphi_1\>$  and $|\varphi_2\>$ we would need an energy
operator $H$ such that
$$
e^{\im t H}|\varphi_1\>  = |\varphi_2\>.
$$
This process would be reversible. (It is worthwhile to note that
the problem of superselection sectors appears also here.)

Von Neumann proved that $S(|\varphi_1\>\<\varphi_1|) \le
S(|\varphi_2\>\<\varphi_2|)$ by constructing a great number of
measurement processes sending the state $|\varphi_1\>$ into an
ensemble, which differs from $|\varphi_2\>\<\varphi_2|$ by an
arbitrarily small amount. The measurement of an observable
$A=\sum_i \lambda_i |\psi_i\>\<\psi_i|$ in a state $|\varphi\>$
yields an ensemble of the pure states $|\psi_i\>\<\psi_i|$ with
weights $|\<\varphi|\psi_i\>|^2$. This was a basic postulate in
von Neumann's measurement theory when the eigenvalues of $A$
are non-degenerate, that is, $\lambda_i$'s are all different.
In a modern language, von Neumann's measurement is a conditional
expectation onto a maximal Abelian subalgebra of the algebra of
all bounded operators acting on the given Hilbert space. Let
$(|\psi_i\>)_i$ be an orthonormal basis consisting of
eigenvectors of the observable under measurement. For any bounded
operator $T$ we set
\begin{equation}
E(T)=\sum_i \<\psi_i|T|\psi_i\>|\psi_i\>\<\psi_i|. \label{8}
\end{equation}
The linear transformation $E$ possesses the following properties:
\begin{itemize}
\item[(i)] $E=E^2$.
\item[(ii)] If $T\ge 0$ then $E(T)\ge 0$.
\item[(iii)] $E(I)=I$.
\item[(iv)] $\tr \big(E(T)\big)=\tr T$.
\end{itemize}

\noindent
In particular, for a statistical operator $D$ its
transform $E(D)$
is a statistical operator as well. It follows immediately from
definition (\ref{8}) that
$$
E(|\varphi\>\<\varphi|)=\sum_i |\<\varphi|\psi_i\>|^2
|\psi_i\>\<\psi_i|
$$
and the conditional expectation $E$ acts on the pure states
exactly in the same way as it is described in the measurement
procedure. It was natural for von Neumann to assume that
\begin{equation}
S(D) \le S\big(E(D)\big),   \label{9}
\end{equation}
at least if the statistical operator $D$ corresponds to a pure
state. Inequality (\ref{9}) is nothing other than the
manifestation of the second law for the measurement process.

In the proof of the inequality $S(|\varphi_1\>\<\varphi_1|) \le
S(|\varphi_2\>\<\varphi_2|)$ one can assume that the vectors
$|\varphi_1\>$ and $|\varphi_2\>$ are orthogonal. The idea is
to construct measurements $E_1,E_2,\dots,E_k$ such that
\begin{equation}
E_k\Big(\dots \big(E_1(|\varphi_1\>\<\varphi_1|)\big)\dots \Big)
\label{10}
\end{equation}
is in a given small neighbourhood of $|\varphi_2\>\<\varphi_2|$.
The details are well-presented in von Neumann's original work,
but we confine ourselves here to his definition for $E_n$. He set
a unit vector
$$
\psi^{(n)}=\cos {\pi n \over 2k}|\varphi_1\>  +
\sin {\pi n \over 2k}|\varphi_2\>
$$
and extended it to a complete orthonormal system. The measurement
conditional expectation $E_n$ corresponds to this basis ($1\le
n\le k$). It is elementary to show that (\ref{10}) tends to
$|\varphi_2\>\<\varphi_2|$ as $k \to \infty$. We stress again
that the argument needs that $|\varphi_1\>$ and $|\varphi_2\>$
are in the same superselection sector, so that their linear combinations
may be state vectors.

Let us summarize von Neumann's discussion of the
thermodynamical entropy of a statistical operator $D$. First of
all, he assumed that $S(D)$ is a continuous function of $D$. He
carried out a reversible process to obtain the mixing property
(\ref{5}) for orthogonal pure states, and he concluded (\ref{6}).
He referred
to the second law again when assuming (\ref{9}) for pure states.
Then he
showed that $S(|\varphi\>\<\varphi|)$ is independent of the state
vector $|\varphi\>$ so that
\begin{equation}
S\Big(\sum_i \lambda_i |\varphi_i\>\<\varphi_i|\Big)=
-\k \sum_i \lambda_i \log \lambda_i  \label{11}
\end{equation}
up to an additive constant which could be chosen to be 0 as a
matter of normalization. (\ref{11}) is von Neumann's celebrated
entropy formula; it has a more elegant form
\begin{equation}
S(D)=\k \tr \eta(D), \label{12}
\end{equation}
where $\eta: \bbbr^+ \to \bbbr$ is the continuous function
$\eta(t)=-t \log t$. (The modern notation for $-t \log t$ comes
from information theory which did not exist at that time.)

When von Neumann deduced (\ref{12}), his natural intention was to
make mild assumptions. For example, the monotonicity (\ref{9}) was
assumed only for pure states. If we already have (\ref{12}) as a
definition, then (\ref{9}) can be proved for an arbitrary statistical
operator
$D$.
The argument is based on the Jensen inequality, and von Neumann
remarked that for
$$
S_f(D)=\tr f(D)
$$
with a differentiable concave function $f:[0,1]\to \bbbr$,
\begin{equation}
S_f(D) \le S_f\big(E(D)\big) \label{13}
\end{equation}
holds for every statistical operator $D$. His analysis also indicated that
the measurement process is typically irreversible, the
finite entropy of a statistical operator definitely increases if
a state change occurs.

Von Neumann solved the maximization problem for $S(D)$ under the
constraint $\tr DH=e$. This means the determination of the
ensemble of maximal entropy when the expectation of the energy
operator $H$ is a prescribed value $e$. It is convenient to
rephrase his argument in terms of conditional expectations.
$H=H^*$ is assumed to have a discrete spectrum and we have a
conditional expectation $E$ determined by the eigenbasis of $H$.
If we pass from an arbitrary statistical operator $D$ with $\tr
DH=e$ to $E(D)$, then the entropy is increasing on the one hand
and the expectation of the energy does not change on the other
hand, so the maximizer should be searched among the operators
commuting with $H$. In this way we are (and von Neumann was) back
to the classical problem of statistical mechanics treated at the
beginning of this article. In terms of operators the solution is
in the form
$$
{\exp (-\beta H) \over \tr \exp (-\beta H)},
$$
which is called the Gibbs state today.

\section{Some topics  about entropy from von Neumann to the
present}

After Boltzmann and von Neumann, it was Shannon who initiated the
interpretation of the quantity $-\sum_i p_i \log p_i$ as ``uncertainty
measure'' or ``information measure''. The American electrical
engineer/scientist Claude Shannon created communication theory in 1948.
He posed a problem in  the following way:

\begin{center}
\parbox[t]{141mm}{
``Suppose we have a set of possible events whose probabilities of
occurence are $p_1,p_2,\dots,p_n$. These probabilities are known but
that is all we know concerning which event will occur. Can we find a
measure of how much ``choice'' is involved in the selection of the event
or how uncertain we are of the outcome?''}
\end{center}

\noindent
Denoting such a measure by $H(p_1,p_2,\dots,p_n)$ he listed three
very reasonable requirements which should be satisfied. He concluded
that the only $H$ satisfying the three assumptions is of the form
$$
H=-K \sum_{i=1}^n p_i\log p_i\,,
$$
where $K$ is a positive constant. For $H$ he used different names such as
information, uncertainty and entropy. Many years later Shannon said
\cite{TM}:

\begin{center}
\parbox[t]{141mm}{
``My greatest concern was what to call it. I thought of
calling it `information', but the word was overly used, so I decided
to call it `uncertainty'. When I discussed it with John von Neumann,
he had a better idea. Von Neumann told me, `You should call it
entropy, for two reasons. In the first place your uncertainty function has
been used in statistical mechanics under that name, so it already has a name.
In the second place, and more important, nobody knows what entropy really is,
so in a debate you will always have the advantage.''}
\end{center}

Shannon's postulates were transformed later into the following axioms:

\begin{itemize}
\item[(a)] Continuity: $H(p,1-p)$ is continuous function of $p$.
\item[(b)] Symmetry: $H(p_1,p_2,\dots ,p_n)$ is a symmetric function
of its variables.
\item[(c)] {Recursion: For every $0\le\lambda <1$ the recursion
$\allowbreak H(p_1,\dots ,p_{n-1},\lambda p_n,(1-\lambda)
p_n)=H(p_1,\dots ,p_n)+p_nH(\lambda,1-\lambda)$ holds.}
\end{itemize}

\noindent
These axioms determine a function $H$ up to a positive
constant factor. Excepting the above story about a conversation
between Shannon and von Neumann, we do not know about any mutual
influence. Shannon was interested in communication theory and 
von Neumann's thermodynamical entropy was in the formalism of quantum
mechanics. Von Neumann himself never made any connection between
his quantum mechanical entropy and information. Although von
Neumann's entropy formula appeared in 1927, there was not much
activity concerning it for several decades. At the end of the 1960's, the
situation changed. Rigorous statistical mechanics came into
being \cite{Ru} and soon after that the needs of rigorous
quantum statistical mechanics forced new developments concerning von
Neumann's entropy formula.

Von Neumann was aware of the fact that statistical operators form
a convex set whose extreme points are exactly the pure states. He
also knew that entropy is a concave functional, so
\begin{equation}
S\Big(\sum_i \lambda_i D_i \Big)\ge \sum_i \lambda S(D_i)
\label{14}
\end{equation}
for any convex combination. To determine the entropy of a
statistical operator, he used the Schatten decomposition, which is
an orthogonal extremal decomposition in our present language. For
a statistical operator $D$ there are many ways to write it in the
form
$$
D=\sum_i \lambda_i |\psi_i\>\<\psi_i|
$$
if we do not require the state vectors to be orthogonal. The
geometry of the statistical operators, that is the state space,
allows many extremal decompositions and among them there is a
unique orthogonal one if the spectrum of $D$ is not degenerate.
Non-orthogonal pure states are essentially nonclassical. They are
between identical and completely different.
Jaynes recognized in 1956 that from the point of view of
information the Schatten decomposition is optimal. He proved that
\begin{eqnarray*}
S(D) & = & \sup \Big\{-\sum_i \lambda_i\log \lambda_i : D=\sum_i
\lambda_i D_i \\
& &  \ \ \ \ \ \ {\rm \ for\ some\ convex\ combination\ and\ statistical\
operators\ }\Big\}.
\end{eqnarray*}
This is Jaynes contribution to the von Neumann entropy \cite{Ja}.
(However, he became known for the very strong advocacy of the
maximum entropy principle.)

Certainly the highlight of quantum entropy theory in the 70's was
the discovery of subadditivity. Before we state it in precise
mathematical form, we describe the setting where this property is
crucial. A one-dimensional quantum lattice system is a composite
system of $2N+1$ subsystems, indexed by the integers
$-N \le n \le N$. Each of the subsystems is described by a
Hilbert space $\iH_n$; those Hilbert spaces are isomorphic if we
assume that the subsystems are physically identical, and even
the very finite dimensional case dim$\iH_n=2$ can be interesting
if the subsystem is a ``spin $1/2$'' attached to the lattice site
$n$. The finite chain of $2N+1$ spins is described in the tensor
product Hilbert space $\otimes_{n=-N}^N \iH_n$, whose dimension is
$(\dim \iH_n)^{2N+1}$. For a given Hamiltonian $H_N$ and inverse
temperature $\beta$ the equilibrium state maximizes the free
energy functional
\begin{equation}
F_N(D_N)= {\rm Tr}_N H_N D_N -{1 \over \beta}S(D_N), \label{15}
\end{equation}
and the actual maximizer is the Gibbs state
\begin{equation}
{\exp (-\beta H_N) \over \tr \exp (-\beta H_N)}. \label{16}
\end{equation}
It seems that this was already known in von Neumann's time but
not the thermodynamical limit, $N\to \infty$. Rigorous statistical
mechanics of spin chains was created in the 70's. Since entropy, energy,
and free energy are extensive quantities, the infinite system should be
handled by their normalized versions, called entropy density, energy
density, etc. One possibility to describe the equilibrium of the
infinite system is to carry out a limiting procedure from the
finite volume equilibrium states, and another is to solve the
variational principle for the free energy density on the state
space of the infinite system. In a translation invariant theory the
two approaches lead to the same conclusion, but many technicalities are
involved. The infinite system is modeled by a $C^*$-algebra and
their states are normalized linear functionals instead of statistical
operators. The rigorous statistical mechanics of quantum spin
systems was one of the successes of the operator algebraic
approach. \cite{Se} and Sec. 15 of \cite{OP} are suggested further readings
about details of quantum spin systems.
One of the key points in this approach is the definition of
entropy density of a state of the infinite system which goes
back to the subadditivity of the von Neumann entropy. Let $\iH_1$
and $\iH_2$ be possibly finite dimensional Hilbert spaces
corresponding to two quantum systems. A mixed state of the
composite system is determined by a statistical operator $D_{12}$
acting on the tensor product $\iH_1 \otimes \iH_2$. Assume that
we are to measure observables on the first subsystem. What is the
statistical operator we need? The statistical operator $D_1$ has
to fulfill the condition
\begin{equation}
{\rm Tr}_1 AD_1 = {\rm Tr}_{12} (A\otimes I)D_{12} \label{18}
\end{equation}
for any observable $A$. Indeed, the left hand side is the
expectation of $A$ in the subsystem and the right hand side is
that in the total system. It is easy to see that condition
\begin{equation}
\<\psi|D_1|\psi\>=\sum_i \< |\psi\>\otimes |\varphi_i\>, D_{12}
|\psi\>\otimes |\varphi_i\>\,\> \label{19}
\end{equation}
gives the statistical operator $D_1$, where $|\psi\>\in \iH_1$
and $|\varphi_i\>$ is an arbitrary orthonormal basis in $\iH_2$.
(In fact equation (\ref{19}) is obtained from (\ref{18}) by putting
$|\psi\>\<\psi|$ in place of $A$.) It is not difficult to state
the subadditivity property now:
\begin{equation}
S(D_{12})\le S(D_1)+S(D_2). \label{20}
\end{equation}
This is a particular case of the strong subadditivity
\begin{equation}
S(D_{123})\le S(D_{12}) + S(D_{23})- S(D_2) \label{21}
\end{equation}
for a system consisting of three subsystems. (We hope that the
notation is selfexplanatory, otherwise see \cite{LR}, \cite{We}
or p. 23 in \cite{OP}.)
If the second subsystem is lacking, (\ref{21}) reduces to
(\ref{20}). (\ref{20})
was proven first by Lieb and Ruskai in 1973 \cite{LR}.

The measurement conditional expectation was introduced by von
Neumann as the basic irreversible state change, and it is of the
form
\begin{equation}
D \mapsto \sum_i P_i D P_i, \label{22}
\end{equation}
where $P_i$ are pairwise orthogonal projections and $\sum_i
P_i=I$. (We are in the Schr\"odinger picture.) The measurement
conditional expectation has the following generalization. Assume
that our quantum system is described by an operator algebra $\iM$
whose positive linear functionals correspond to the states. A
functional $\tau : \iM \to \bbbc$ is a state if $\tau(A) \ge 0$
for any positive observable $A$ and $\tau(I)=1$. An operational
partition of unity is a finite subset $\iV=(V_1,V_2,\dots,V_n)$ of
$\iM$ such that $\sum_i V_i^*V_i=I$. In the Heisenberg picture
$\iV$ acts on the observables as
$$
A \mapsto \sum_i V_i^* A V_i
$$
and the corresponding state change in the Schr\"odinger picture
is
$$
\tau(\pont)\mapsto \tau\Big(\sum_iV_i^* \pont V_i\Big).
$$
Let us compare this with the traditional formalism of quantum
mechanics. If $\tau(A)=\tr DA$, then
$$
\tau\Big(\sum_iV_i^* A V_i\Big)=\tr \Big(D\sum_iV_i^* A V_i\Big)=
\tr \Big(\sum_iV_i D V_i^*\Big)A,
$$
hence the transformation of the statistical operator is
$$
D \mapsto \sum_iV_i D V_i^*
$$
which is an extension of von Neumann's measurement conditional
expectation (\ref{22}). Given a state $\tau$ of the quantum system, the
observed entropy of the operational process is defined to be the
von Neumann entropy of the finite statistical operator
$$
\big[\tau(V_i^*V_j)\big]_{i=j=1}^n,
$$
which is an $n\times n$ positive semidefinite matrix of trace 1.
If we are interested in the entropy of a state, we perform all
operational processes and compute their entropy. If the
operational process changes the state of our system, then the
observed operational entropy includes the entropy of the state
change. Hence we have to restrict ourself to state invariant
operational processes when focusing on the entropy of the state.
The formal definition
$$
S^L(\tau)=\sup\Big\{S\big(\big[\tau(V_i^*V_j)\big]_{i=j=1}^n\big)\Big\}
$$
is the operational (or Lindblad) entropy of the state $\tau$ if
the least upper bound is taken over all operational partitions of unity
$\iV=(V_1,V_2,\dots,V_n)$ such that
$$
\tau(A)\mapsto \tau\Big(\sum_iV_i^* A V_i\Big),
$$
for every observable $A$. For a statistical operator $D$ we have
$$
S^L(D)=2 S(D),
$$
and we may imagine that the factor 2 is removable by appropriate
normalization, so that we are back to the von Neumann entropy. The
operational entropy satisfies von Neumann's mixing condition and
is a concave functional on the states even in the presence of
superselection rules. However, it has some new features. To see
a concrete example, assume that there are two superselection
sectors and the operator algebra is $M_2(\bbbc)\oplus
M_3(\bbbc)$, that is, the direct sum of two full matrix algebras.
Let a state $\tau_0$ be the mixture of the orthogonal pure states
$|\psi_i\>$ with weights $\lambda_i$, where $|\psi_1\>,|\psi_2\>$
are in the first sector and $|\psi_3\>$ is
in the second. This assumption implies that there is no dynamical
change sending $|\psi_1\>$ into $|\psi_3\>$, and superpositions
of those states are also prohibited. One computes
\begin{eqnarray*}
S^L(\tau_0) & = &-2\sum_i \lambda_i \log \lambda_i \\
& & - (\lambda_1+\lambda_2) \log (\lambda_1+\lambda_2)-
(\lambda_1+\lambda_2+\lambda_3) \log
(\lambda_1+\lambda_2+\lambda_3),
\end{eqnarray*}
which shows that this entropy is really sensitive to the
superselection sectors. (For further properties on $S^L$ we refer
to pp. 121--124 of \cite{OP}.)

Nowdays some devices are based on quantum mechanical phenomena,
and this holds also for information transmission. For example, in
optical communication a polarized photon can carry information.
Although von Neumann apparently did not see an intimate connection
between his entropy formula and the formally rather similar
Shannon information measure, many years later an information
theoretical reinterpretation of von Neumann's entropy is becoming
common. Communication theory deals with coding, transmission, and
decoding of messages. Given a set $\{a_1, a_2,\dots, a_n\}$ of
messages, a coding procedure assigns to each $a_i$ a physical
state, say a quantum mechanical state $|\psi_i\>$. The states are
transmitted and received. During the transmission some noise can
enter. The receiver uses some observables to recover the
transmitted message. Shannon's classical model is stochastic, so it
is assumed that each message $a_i$ should be teleported with some
probability $\lambda_i$, $\sum_i \lambda_i=1$. Hence in the
quantum model the input state of the channeling transformation is a
mixture; its statistical operator is $D_{{\rm in}}=\sum_i p_i
|\psi_i\>\<\psi_i|$.
This is the state we need to transmit, and after transmission it
changes into $T(D_{{\rm in}})=D_{{\rm out}}$ which is
formally a
statistical operator but may correspond to a state of a very
different system. Input and output could be far away in space as
well as in time. The observer measures the observable $A_j$ and
$p_i=\tr
D_{{\rm out}}A_i$ is the probability with which he concludes the
message $a_j$ was transmitted. Here we need $\sum_j A_j=I$ and
$0\le A_i$. More generally, we assume that
$$
p_{ji}=\tr {T}(|\psi_i\>\<\psi_i|) A_j
$$
is the probability that the receiver deduces the transmission of
the message $a_j$ when actually the message $a_i$ was
transmitted. If we forget about the quantum mechanical coding,
transmission and decoding (measurement), we see a classical
information channel in Shannon's sense. According to Shannon,
the amount of information going through the channel is
$$
I=\sum_{ij} p_{ji} \log {p_{ji} \over p_i}.
$$
One of the basic problems of communication theory is to maximize
this quantity subject to certain constraints. For the sake of
simplicity, assume that there is no noise. This may happen when
the channel is actually the memory of a computer; storage of
information might be a noiseless channel in Shannon's sense.
We have then ${T}=$identity, $D_{{\rm in}}=D_{{\rm out}}=D$
and the inequality
$$
I\le S(D)
$$
holds. If we fix the channel state $D$ and optimize with respect
to the probabilities $\lambda_i$, the states $|\psi_i\>$ and the
observables $A_j$, then the maximum information
transmittable through the channel is exactly the von Neumann
entropy. What we are considering is a simple example, probably
the simplest possible. However, it is well demonstrated that the
von Neumann entropy is actually the capacity of a communication
channel. Recently, there has been a lot of discussion about
capacities of quantum communication channels, which is outside of
the scope of the present article. However, the fact that von
Neumann's entropy formula has much to do with Shannon theory and
possesses an interpretation as measure of information must be
conceptually clear without entering more sophisticated models and
discussions. More details are in \cite{Pe} and a mathematically full 
account is \cite{Ho}.

Further sources about quantum entropy and quantum information are
\cite{Jo}, \cite{Petz} and \cite{OP}.

\end{document}